\documentclass[graybox]{svmult}

\usepackage{type1cm}        

\usepackage{makeidx}         
\usepackage{graphicx}        
\usepackage{multicol}        
\usepackage[bottom]{footmisc}

\usepackage{newtxtext}       %
\usepackage{newtxmath}       
\UseRawInputEncoding

\makeindex             

\begin{document}

\title*{Probing the horizon of black holes with gravitational waves}
\author{Elisa Maggio}
\institute{Elisa Maggio \at Max Planck Institute for Gravitational Physics, Albert Einstein Institute,  Am M\"uhlenberg 1, 14476 Potsdam, Germany\\ \email{elisa.maggio@aei.mpg.de}
}

\maketitle

\abstract{
Gravitational waves open the possibility to investigate the nature of compact objects and probe the horizons of black holes. Some models of modified gravity predict the presence of horizonless and singularity-free compact objects.
Such dark compact objects would emit a gravitational-wave signal which differs from the standard black hole scenario.
In this chapter, we overview the phenomenology of dark compact objects by analysing their characteristic frequencies in the ringdown and the emission of gravitational-wave echoes in the postmerger signal. We show that future gravitational-wave detectors will allow us to perform model-independent tests of the black hole paradigm.
}

\section{Tests of the black hole paradigm}

Black holes (BHs) are the end result of the gravitational collapse and the most compact objects in the Universe. According to the no-hair theorems of general relativity (GR), any compact object heavier than a few solar masses is well described by the Kerr geometry~\cite{Carter:1971zc,Robinson:1975bv}. Kerr BHs are determined uniquely by two parameters, i.e., their mass $M$ and angular momentum $J$ defined through the dimensionless spin parameter $\chi \equiv J/M^2$~\cite{Kerr:1963ud}. Therefore, any observation of deviation from the properties of Kerr BHs would be an indication of departure from GR. 

Gravitational waves (GWs) provide a unique channel for probing the nature of astrophysical sources. The GW signal emitted by the coalescence of compact binaries is characterized by three main stages: the \emph{inspiral}, when the two bodies spiral in towards each other as they loose energy into gravitational radiation; the \emph{merger}, when the two bodies coalesce; and the \emph{ringdown}, when the final remnant relaxes to an equilibrium solution. 
In particular, the analysis of the ringdown would allow us to infer the properties of the compact remnants. 

The ringdown is dominated by the complex characteristic frequencies of the remnant, the so-called \emph{quasi-normal modes} (QNMs), which describe the response of the compact object to a perturbation~\cite{Chandrasekhar:1975zza}, i.e.
\begin{equation}
\omega_{\ell m n} = \omega_{R, \ell m n} + i \omega_{I, \ell m n} \,,
\end{equation}
where $\omega_{R/I, \ell m \omega} \in \Re$. Each mode is described by three integers, namely the angular number of the perturbation $\ell$ (where $\ell \geq 0$), the azimuthal number of the perturbation $m$ (such that $|m| \leq \ell$), and the overtone number $n$ (where $n \geq 0$). The fundamental mode with $n = 0$ corresponds to the mode with the smallest imaginary part.
The ringdown is modeled as a sum of exponentially damped sinusoids whose frequencies $f_{\ell m n}$ (damping times $\tau_{\ell m n}$) are related to the real (imaginary) part of the QNMs of the remnant via
\begin{eqnarray}
f_{\ell m n} &=& \omega_{R, \ell m n}/(2 \pi) \,, \\
\tau_{\ell m n} &=& -1/\omega_{I, \ell m n} \,.
\end{eqnarray}
Therefore, from the detection of the ringdown signal it is possible to infer the QNMs of the remnant and understand the nature of the latter.

The fundamental QNM has been observed in the ringdown of several GW events~\cite{LIGOScientific:2021sio}. The ringdown detections are compatible with Kerr BH remnants, however the characterization of the remnant requires further analyses. Indeed, the measurement of one complex QNM allows us only to estimate the mass and the spin of the remnant. A test of the BH paradigm would require the identification of at least two QNMs in the ringdown. 
Next generation detectors, e.g. the space-based interferometer LISA, will allow for tests of the BH paradigm with unprecedented precision~\cite{Berti:2005ys}.
 
\section{Horizonless compact objects}

On the theoretical side, the presence of horizons in Kerr BHs poses some issues.
In particular, the horizon hides a curvature singularity with infinite tidal forces where the Einstein equations break down.
Moreover, the spacetime within the horizon can contain closed time-like hypersurfaces that violate causality. 

Several attempts to regularize the BH solution predict the existence of horizonless and singularity-free compact objects~\cite{Cardoso:2019rvt}. Some models are solutions to quantum-gravity extensions of GR, e.g. the fuzzball in string theory as an ensemble of a large number of regular and horizonless microstate geometries with the same asymptotic charges of a BH~\cite{Mathur:2005zp}. Other models of horizonless compact objects are solutions to GR in the presence of dark matter or exotic fields, e.g. boson stars as self-gravitating solutions formed by massive bosonic fields which are coupled minimally to GR~\cite{Liebling:2012fv}. 

Horizonless compact objects can mimic BHs in terms of electromagnetic observations since they can be as compact as BHs~\cite{Abramowicz:2002vt}. For example, the observation of the supermassive object at the center of the galaxy M87 by the Event Horizon Telescope constrained weakly some models of horizonless compact objects~\cite{EventHorizonTelescope:2019pgp}. Moreover, horizonless compact objects can be used to study GW events in the mass gap between neutron stars and BHs and due to pair-instability supernova processes~\cite{Bustillo:2020syj,LIGOScientific:2020zkf}.

In this context, horizonless compact objects allow us to quantify the existence of horizons in astrophysical sources. We analyse a generic model of \emph{dark compact object} which deviate from a BH for two parameters~\cite{Maggio:2021ans}:
\begin{itemize}
\item the compactness, which is defined as the inverse of the effective radius of the object in units of mass, i.e. $C = M/r_0$,
where
\begin{equation}\label{r0}
r_0 = r_+ (1+\epsilon)
\end{equation}
is the location of the effective radius of the object and $r_+ = M \left(1+\sqrt{1-\chi^2}\right)$ is the horizon of a Kerr BH.
Depending on their compactness, two categories of horizonless compact objects can be distinguished: compact objects whose effective radius is comparable with the light ring of BHs, i.e. $\epsilon \approx 0.1,1$; and ultracompact objects with Planckian corrections at the horizon scale due to quantum fluctuations, i.e. $\epsilon \approx 10^{-40}$. The two categories of horizonless compact objects give rise to different fingerprints in the GW signal. In particular, a merger remnant with $\epsilon \approx 0.1,1$ would emit a ringdown signal which differs from the BH ringdown at early stages, whereas an ultracompact horizonless object would emit a modulated train of GW echoes at late times, as discussed in Sec.~\ref{sec:echoes};
\item the ``darkness'', which is related to the reflectivity of the compact object $\mathcal{R}(\omega)$ at its effective radius. The BH is a totally absorbing object with $\mathcal{R}=0$ at the horizon, whereas a horizonless compact object can have $0 \leq |\mathcal{R}(\omega)|^2 \leq 1$ depending on its interior structure. The $|\mathcal{R}(\omega)|^2 = 1$ case describes a perfectly reflecting object of perturbations moving towards the object. This is the case, for example, of neutron stars where the absorption of radiation through viscosity is negligible. Intermediate values of $\mathcal{R}(\omega)$ describe partially absorbing compact objects due to dissipation, viscosity, fluid mode excitations, nonlinear effects, etc.

\end{itemize} 

\section{Phenomenology}

Let us derive the GW signatures of horizonless compact objects in the postmerger phase of compact binary coalescences. In this section, we overview the quasi-normal mode spectrum and the GW signal in the time domain at variance with the BH case.

\subsection{Quasi-normal mode spectrum} \label{sec:QNMs}

\begin{figure}[t]
\centering
\includegraphics[width=0.7\textwidth]{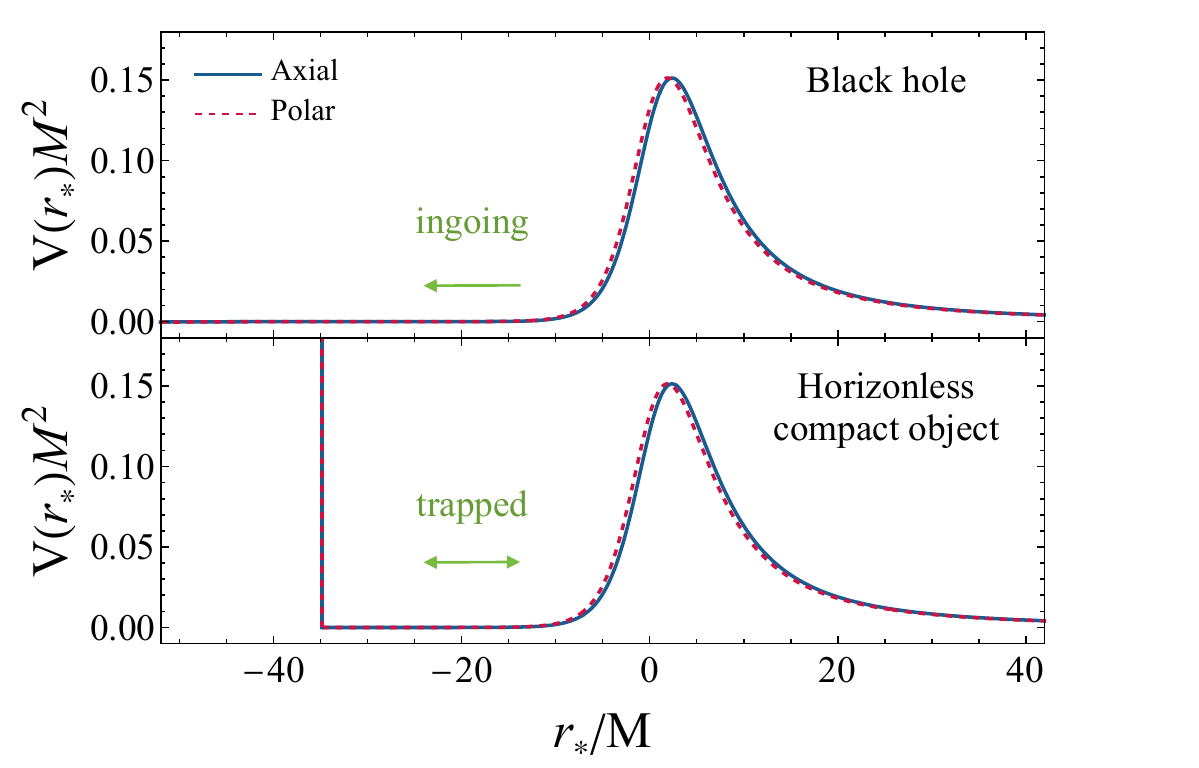}
\caption{Effective potential as a function of the tortoise coordinate of a Schwarzschild BH (top panel) and a static horizonless  compact object  with radius $r_0 = 2M(1+\epsilon)$ (bottom panel), for axial (continuous line) and polar (dashed line) $\ell=2$ gravitational perturbations. The effective potential has a peak approximately at the light ring, $r \approx 3M$. In the case of a horizonless compact object, the effective potential features a cavity between the radius of the object and the light ring. Adapted from~\cite{Cardoso:2016rao,Cardoso:2019rvt,Maggio:2021ans}.} 
\label{fig:potential}
\end{figure}
For simplicity, let us analyse a static and spherically symmetric horizonless compact object. Let us assume that GR is a reliable approximation outside the radius of the object and some modifications appear at the horizon scale. Owing to the Birkhoff theorem, the exterior spacetime is described by the Schwarzschild metric
\begin{equation}\label{schw}
ds^2 = -f(r) dt^2 + \frac{1}{f(r)} dr^2 + r^2 \left( d\theta^2 + \sin^2 \theta d\phi^2\right) \,, 
\end{equation}
where $(t,r,\theta,\phi)$ are the Boyer-Lindquist coordinates and $f(r)=1-2M/r$. The radius of the compact object is located as in Eq.~\eqref{r0}, where $r_+ = 2M$ is the horizon of a Schwarzschild BH. In order to derive the QNM spectrum of the horizonless compact object, let us perturb the background geometry with a gravitational perturbation. The radial component of the gravitational perturbation is governed by a second-order differential equation~\cite{Regge:1957td,Zerilli:1970se}
\begin{equation}\label{eqdiff}
 \frac{d^2 \psi(r)}{dr_*^2} + \left[ \omega^2 - V(r)\right] \psi(r) = 0 \,,
\end{equation}
where $r_*$ is the tortoise coordinate defined such that $dr_*/dr = 1/f(r)$, $f(r)$ is the Schwarzschild function $f(r)=1-2M/r$, and the effective potential reads
\begin{eqnarray}
    V_{\rm axial}(r) &=& f(r) \left[ \frac{\ell(\ell+1)}{r^2} - \frac{6M}{r^3}\right] \,, \label{Vaxial} \\
    V_{\rm polar}(r) &=& 2f(r) \left[ \frac{q^2 (q+1) r^3 + 3 q^2  M r^2 + 9 M^2 (q r + M)}{r^3 (qr+3M)^2} \right] \,, \label{Vpolar}
\end{eqnarray}
for axial and polar perturbations, respectively, with parity $(-1)^{\ell+1}$ and $(-1)^{\ell}$, where $q=(\ell-1)(\ell+2)/2$. Fig.~\ref{fig:potential} shows the effective potential as a function of the tortoise coordinate for a BH (top panel) and a horizonless compact object (bottom panel). The effective potentials display a peak approximately at the light ring, $r \approx 3M$, which is the unstable circular orbit of photons around the compact object. In the BH case, the perturbation is purely ingoing towards the horizon; whereas in the case of a horizonless compact object, the absence of the horizon implies the existence of a cavity between the radius of the object and the light ring. The cavity can support trapped modes that are responsible for a completely different QNM spectrum with respect to the BH case.

By adding two boundary conditions to Eq.~\eqref{eqdiff}, the system defines an eigenvalue problem whose complex eigenvalues are the QNMs of the object.
At infinity, we impose that the perturbation is a purely outgoing wave, i.e.
\begin{equation}
    \psi(r) \sim e^{i \omega r_*} \,, \quad \text{as} \ r_* \to + \infty \,. \label{infBC}
\end{equation}
In the case of a horizonless ultracompact object ($\epsilon \ll 1$), the perturbation can be decomposed a superposition of ingoing and outgoing waves at the radius of the object, i.e.
\begin{equation}
    \psi(r) \sim C_{\rm in}(\omega) e^{-i \omega r_*} + C_{\rm out}(\omega) e^{i \omega r_*} \,, \quad \text{as} \ r_* \to r_*^0 \,, \label{solr0}
\end{equation}
where the reflectivity of the compact object is defined as~\cite{Maggio:2017ivp}
\begin{equation}
    \mathcal{R}(\omega) = \frac{C_{\rm out}(\omega)}{C_{\rm in}(\omega)} e^{2 i \omega r_*^0} \,. \label{R}
\end{equation}
\begin{figure}[t]
\centering
\includegraphics[width=0.7\textwidth]{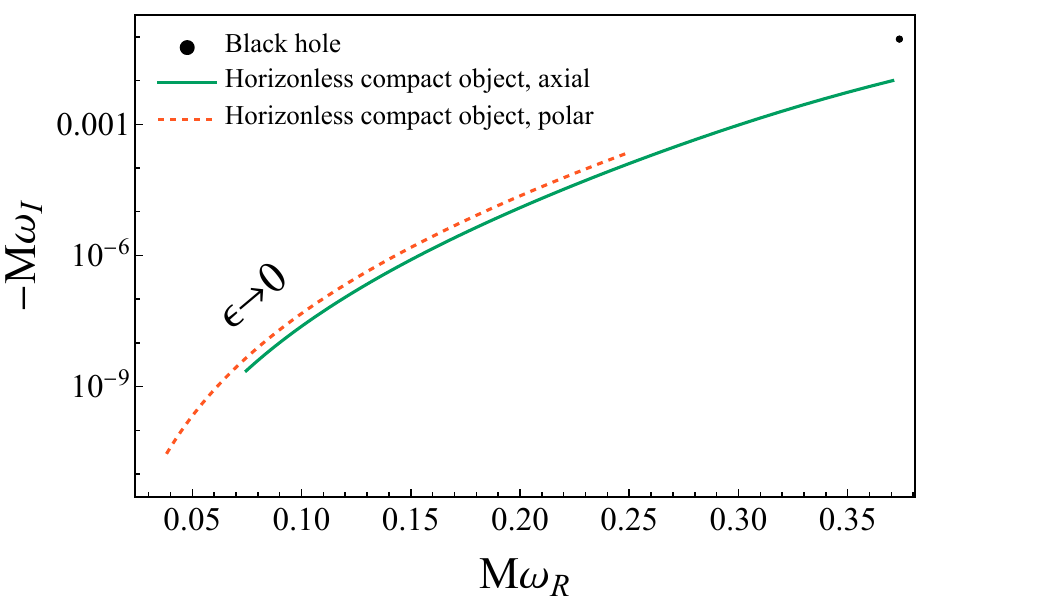}
\caption{QNM spectrum of a perfectly reflecting horizonless compact object with radius $r_0=2M(1+\epsilon)$ and $\epsilon \in (10^{-10},10^{-2})$ compared to the fundamental $\ell=2$ QNM of a Schwarzschild BH. Axial and polar modes are not isospectral as in the BH case. As $\epsilon \to 0$, the QNM spectrum is low-frequencies and long-lived. Adapted from~\cite{Cardoso:2016rao,Maggio:2021ans}.}
\label{fig:lowfrequencyQNMs}
\end{figure}
Let us derive the fundamental ($n=0$) $\ell=2$ QNM which is the mode with the longest damping time (in the static and spherically symmetric case, the QNMs do not depend on the azimuthal number $m$). Fig.~\ref{fig:lowfrequencyQNMs} shows the QNM spectrum of a horizonless ultracompact object with a perfectly reflecting surface $\left(|\mathcal{R}(\omega)|^2=1 \right)$ and $\epsilon \in \left(10^{-10},10^{-2} \right)$ from the left to the right of the plot compared to the fundamental $\ell=2$ QNM of a Schwarzschild BH, i.e.
\begin{equation}
M \omega_{\rm{BH}} = 0.3737 - i 0.08896 \,.
\end{equation}
A first important signature of horizonless compact object is the breaking of isospectrality between axial and polar modes differently from BHs in GR. Indeed, Schwarzschild BHs have a unique QNM spectrum~\cite{Chandrasekhar:1975zza} despite the effective potentials for axial and polar perturbations differ from each other  (see Eqs.~\eqref{Vaxial},~\eqref{Vpolar}). Conversely, the radius of horizonless compact objects is responsible for the appearance of a mode doublet for axial and polar QNMs.
\vspace{0.5cm}
\hrule
\subsubsection*{Exercise}
The isospectrality of axial and polar modes in BHs can be demonstrated from the Darboux transformation between the Regge-Wheeler and Zerilli wave functions governing axial and polar modes, respectively, both satisfying Eq.~\eqref{eqdiff}, i.e.
\begin{equation}\label{darboux}
	\psi_{\text{RW}} = A \frac{d \psi_{\text{Z}}}{dr_*} + B(r) \psi_{\text{Z}} \,,
\end{equation}
where 
\begin{eqnarray}
	A &=& - M \left[ i \omega M + \frac{1}{3} q(q+1)\right]^{-1} \,, \\
	B(r) &=& \frac{q(q+1)(q r + 3 M) r^2 + 9 M^2 (r-2M)}{r^2 (q r +3M)[q(q+1)+3 i \omega M]}  \,.
\end{eqnarray}
Demonstrate that the BH boundary condition $\psi = C_{\text{in}(\omega)} e^{-i \omega r_*}$ as $r \to 2M$
for both Regge-Wheeler and Zerilli wave functions satisfies the Darboux transformation in Eq.~\eqref{darboux}. Conversely, demonstrate that the boundary condition of a horizonless ultracompact object in Eq.~\eqref{solr0} does not satisfy the Darboux transformation in Eq.~\eqref{darboux}.
\vspace{0.5cm}
\hrule
Furthermore, a relevant feature of horizonless compact objects is that the QNM spectrum is low-frequency and long-lived in the limit $\epsilon \to 0$. 
For example, the fundamental $\ell=2$ QNMs of a perfectly reflecting compact object with $\epsilon=10^{-10}$ are:
\begin{eqnarray}
M \omega_{\rm axial} = 0.07470 - i 2.299 \times 10^{-9} \,, \\
M \omega_{\rm polar} = 0.03791 - i 2.739 \times 10^{-11} \,.
\end{eqnarray}
This finding might seem surprising since, in the limit of a compactness close to the BH case, the QNM spectrum of a horizonless compact object deviates significantly from the BH QNM spectrum. A key role is played by the boundary condition in Eq.~\eqref{solr0}, particularly by the fact that the reflective properties of a horizonless compact object differ generically from the totally absorbing BH case.  

Low-frequency QNMs can be understood in terms of the trapped modes between the  radius of the compact object and the light ring, as shown in Fig.~\ref{fig:potential}.
The real part of the QNMs depends on the width of the cavity in the effective potential, whereas the imaginary part of the QNMs depends on the amplification factor of the modes in the cavity and the reflectivity at the radius of the compact object.
For $\epsilon \ll 1$, the QNMs can be derived analytically in the low-frequency regime as~\cite{Maggio:2018ivz,Cardoso:2019rvt,Vilenkin:1978uc,Starobinskil:1974nkd}
\begin{eqnarray}
 \omega_R &\sim& - \frac{\pi}{2 |r_*^0|} \left(p + 1\right) \,, \label{MomegaRa0} \\
 \omega_I &\sim& - \frac{\beta_{2 \ell}}{|r_*^0|} \left( 2 M \omega_R\right)^{2 \ell +2} \,, \label{MomegaIa0}
\end{eqnarray}
where $\sqrt{\beta_{2 \ell}} = \frac{(\ell-2)! (\ell+2)!}{(2\ell)! (2\ell+1)!!}$ and $p$ is a positive odd (even) integer for polar (axial) modes. The real part of the QNMs scales with the compactness of the object as $\omega_R \sim |\log \epsilon|^{-1}$, whereas the imaginary part of the QNMs scales as $\omega_I \sim -|\log \epsilon|^{-(2 \ell +3)}$.

Let us notice that the boundary condition in Eq.~\eqref{solr0} can be imposed at the radius of the compact object when $\epsilon \ll 1$ and the effective potential is vanishing. To derive the QNMs of horizonless compact objects with any compactness, we can make use of the membrane paradigm.
The original BHs membrane paradigm states that a static observer outside the BH horizon can replace the interior of the perturbed BH by a \emph{fictitious} membrane located at the horizon~\cite{damour,membrane}. 
The generalisation of the membrane paradigm to horizonless compact objects allows us to describe any compact object with a Schwarzschild exterior where no specific model is assumed for the object's interior. The compactness of the horizonless object is generic and the reflectivity of the object is mapped in terms of the properties of the fictitious membrane.

The Israel-Darmois junction conditions fix the properties of the fictious membrane relating the exterior and the interior spacetime to the radius of the compact object, i.e.~\cite{darmois,Israel:1966rt}
\begin{equation} \label{junction}
    [[K_{ab} - K h_{ab}]]=-8 \pi T_{ab} \,, \qquad [[h_{ab}]]=0 \,,
\end{equation}
where $h_{ab}$ is the induced metric on the membrane, $K_{ab}$ is the extrinsic curvature, $K = K_{ab}h^{ab}$, $T_{ab}$ is the membrane stress-energy tensor, and $[[...]]$ is the jump of a quantity across the membrane (detailed definitions of the above quantities are in Ref.~\cite{Maggio:2020jml}). For the membrane paradigm, the fictitious membrane is such that the extrinsic curvature of the interior spacetime vanishes. As a consequence, Eqs.~\eqref{junction} impose that the fictitious membrane is a \emph{viscous} fluid with stress-energy tensor 
\begin{equation} \label{stressenergytensor}
    T_{ab} = \rho u_a u_b + (p- \zeta \Theta) \gamma_{ab} -2 \eta \sigma_{ab} \,,
\end{equation}
where $\eta$ and $\zeta$ are the shear and bulk viscosities of the fluid, $\rho$, $p$ and $u_a$ are the density, pressure and 3-velocity of the fluid, $\Theta=u^a_{;a}$ is the expansion, $\sigma_{ab}$ is the shear tensor, and the semicolon is the covariant derivative compatible with the induced metric.
BHs are described by the following values of the shear and bulk viscosities of the membrane:
\begin{equation}\label{etazetaBH}
    \eta_{\rm BH} = \frac{1}{16 \pi} \,, \quad \zeta_{\rm BH} = -\frac{1}{16 \pi} \,;
\end{equation}
whereas horizonless compact objects have values of the shear and bulk viscosities which are generically complex and frequency dependent. For a specific model for the interior of the compact object, the shear and the bulk viscosities are uniquely determined.
The junction conditions in Eq.~\eqref{junction} with the stress-energy tensor in Eq.~\eqref{stressenergytensor} allow us to derive the boundary conditions at the radius of the horizonless compact object, i.e.~\cite{Maggio:2020jml}
\begin{eqnarray}
\frac{d \psi(r_0)/dr_*}{\psi(r_0)} &=& - \frac{i \omega}{16 \pi \eta} - \frac{r_0^2 V_{\rm axial}(r_0)}{2(r_0-3M)} \,, \qquad \text{axial} \,, \label{BC-axial}\\
\frac{d \psi(r_0)/dr_*}{\psi(r_0)} &=& - 16 \pi i \eta \omega + G(r_0,\omega,\eta,\zeta) \,, \quad \text{polar} \,, \label{BC-polar}
\end{eqnarray}
where $G(r_0,\omega,\eta,\zeta)$ is a cumbersome function given in Ref.~\cite{Maggio:2020jml}.
The boundary conditions in Eqs.~\eqref{BC-axial},~\eqref{BC-polar} describe a horizonless object with any compactness whose reflective properties are mapped in terms of the shear and bulk viscosities of the fictitious membrane.
\vspace{0.5cm}
\hrule
\subsubsection*{Exercise}
\begin{itemize}
\item[1.] Demonstrate that, in the limit ($r_0 \to 2M$), the axial boundary condition in Eq.~\eqref{BC-axial} reduces to a purely ingoing wave when the condition in Eq.~\eqref{etazetaBH} is satisfied.
\item[2.] For $\epsilon \ll 1$, the axial boundary condition in Eq.~\eqref{BC-axial} reduces to the boundary condition in Eq.~\eqref{solr0} for horizonless ultracompact objects. Derive that the relation between the reflectivity of the compact object and the shear viscosity of the membrane is in the large-frequency limit:
\begin{equation}
	|\mathcal{R}|^2 = \left( \frac{1- \eta / \eta_{\text{BH}}}{1+ \eta / \eta_{\text{BH}}} \right)^2 \,.
\end{equation}
This shows that a compact object is a perfect absorber of high-frequency waves ($|\mathcal{R}|^2=0$) if $\eta = \eta_{\rm BH}$, whereas it is a perfect reflector of high-frequency waves ($|\mathcal{R}|^2=1$) when either $\eta=0$ or $\eta \to \infty$.
\end{itemize}
\vspace{0.5cm}
\hrule
\vspace{0.5cm}
\begin{figure}[t]
\centering
\includegraphics[width=0.49\textwidth]{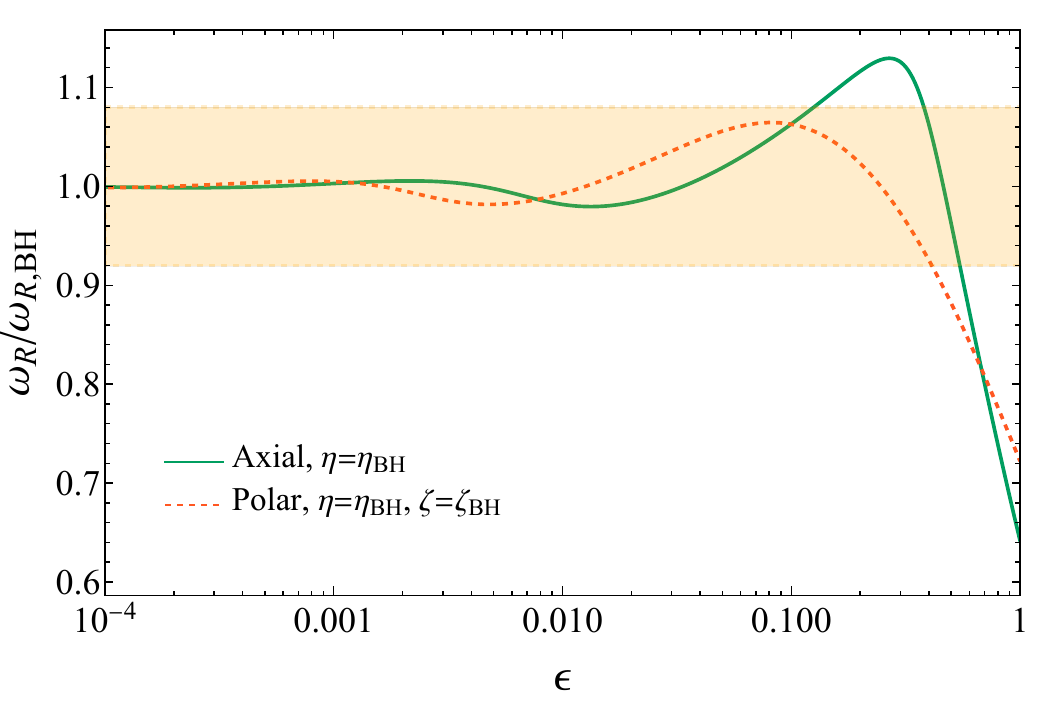}
\includegraphics[width=0.49\textwidth]{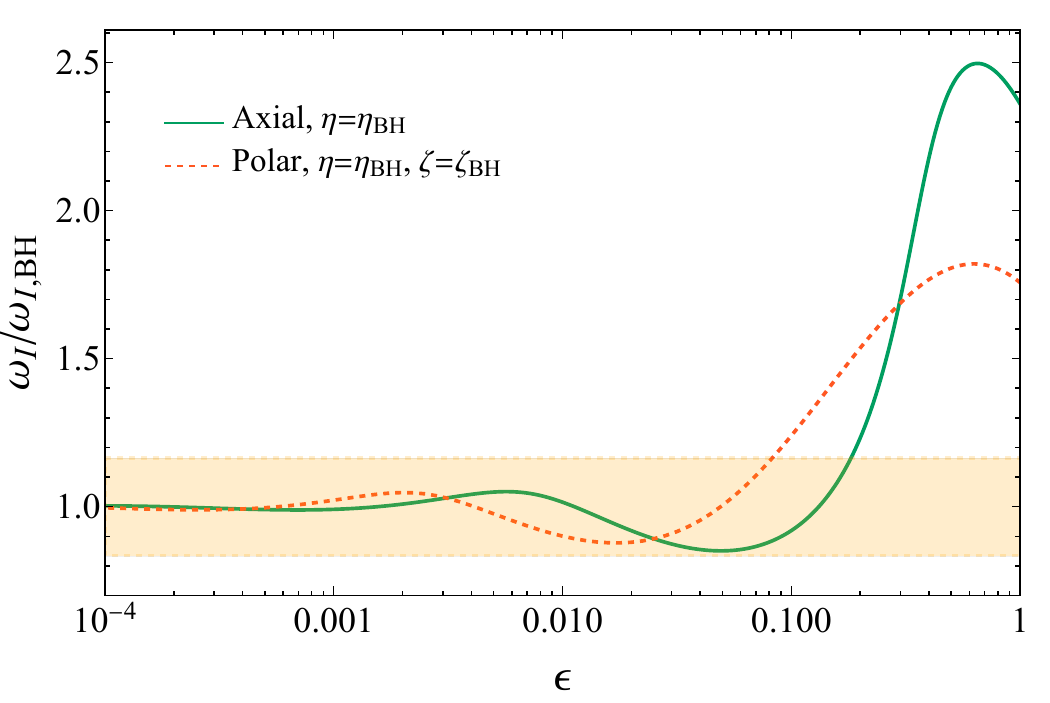}
\caption{Real (left panel) and imaginary (right panel) part of the QNMs of a horizonless compact object described by a fictitious fluid with shear viscosity $\eta=\eta_{\rm BH}$ and bulk viscosity $\zeta=\zeta_{\rm BH}$ compared to the fundamental $\ell=2$ QNM of a Schwarzschild BH, as a function of $\epsilon$ where the radius of the object is located at $r_0 = 2M(1+\epsilon$). The highlighted region is the maximum deviation (with $90\%$ credibility) for the least-damped QNM in the event GW150914~\cite{Ghosh:2021mrv}. Horizonless compact objects with $\epsilon \lesssim 0.1$ are compatible with current measurement accuracies. Adapted from~\cite{Maggio:2020jml,Maggio:2021ans}.}
\label{fig:QNMsmembrane}
\end{figure}

Fig.~\ref{fig:QNMsmembrane} shows the ratio of the real (left panel) and imaginary (right panel) part of the QNMs of a horizonless compact object to the fundamental $\ell=2$ QNM of a Schwarzschild BH as a function of the compactness. Let us notice that as $\epsilon \to 0$, the QNM spectrum of the horizonless compact object coincides with the BH spectrum. This is because a horizonless compact object with the shear and bulk viscosities as in Eq.~\eqref{etazetaBH} has the same reflective properties of a BH.
For relatively large values of $\epsilon$, the compactness of the object decreases and the QNMs deviate from the BH QNM. 
The highlighted regions are the maximum allowed deviation (with $90\%$ credibility) for the least-damped QNM in the event GW150914, and correspond to $\sim 16\%$ and $\sim 33\%$ for the real and imaginary part of the QNM, respectively~\cite{Ghosh:2021mrv}. Fig.~\ref{fig:QNMsmembrane} shows that horizonless compact objects with $\epsilon \lesssim 0.1$ are compatible with current measurement accuracies. Next-generation detectors would allow us to set more stringent constraints on the radius of compact objects.

\subsection{Gravitational-waves echoes} \label{sec:echoes}

In this section, we shall analyse the modifications that would appear in the postmerger GW signal if the remnant of a compact binary coalescence is a horizonless compact object. The phenomenology depends strongly on the compactness of the object. In particular, if the remnant is a horizonless ultracompact object ($\epsilon \ll 1$) the prompt ringdown would be nearly indistinguishable from the BH ringdown since it is due to the excitation of the light ring 
that occurs approximately at the same location as shown in Fig.~\ref{fig:potential}. Afterwards, some trapped modes travel within the cavity of the effective potential and are reflected back at the radius of the compact object. After the interaction with the light ring, an additional GW signal is emitted at infinity in the form a GW echo. Multiple reflections of the trapped modes in the cavity can give rise to a train of GW echoes.

\begin{figure}[t]
\centering
\includegraphics[width=0.49\textwidth]{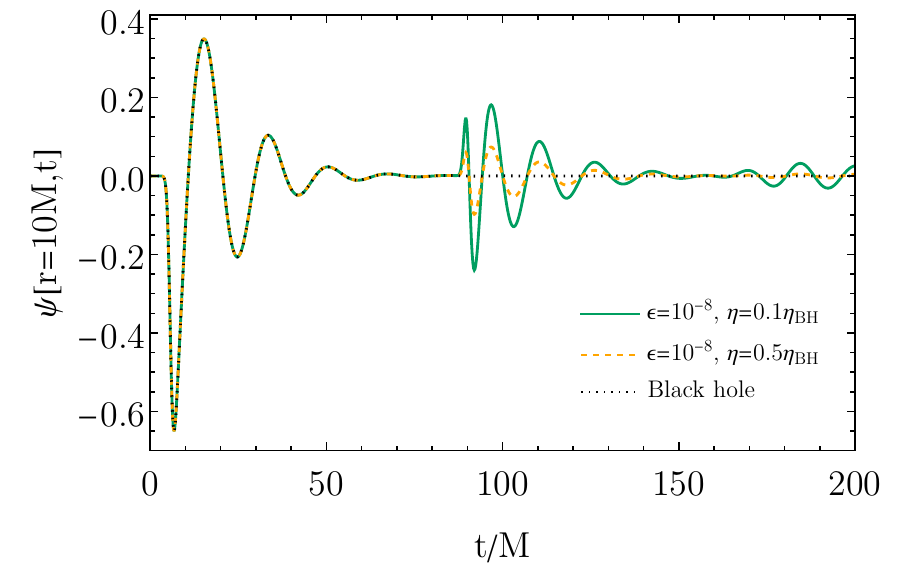}
\includegraphics[width=0.49\textwidth]{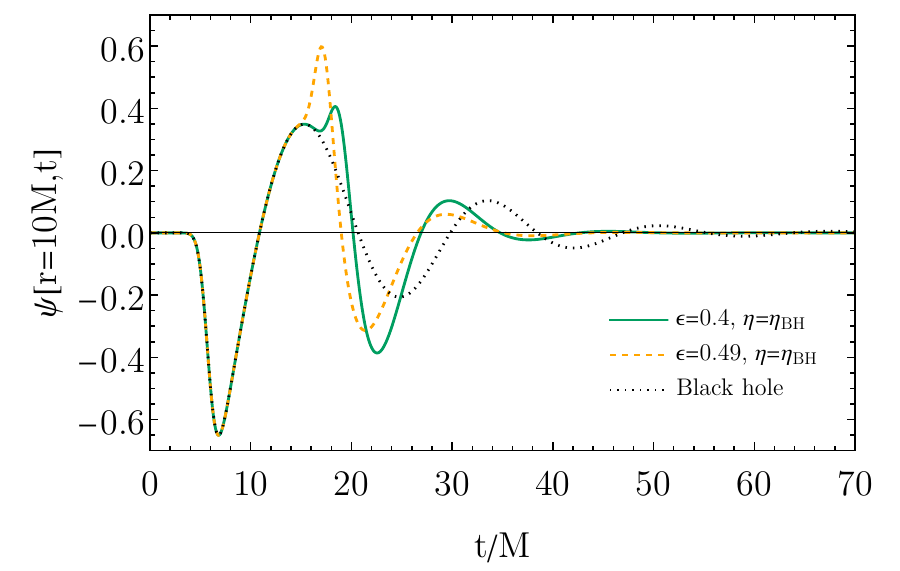}
\caption{Left panel: GW echoes emitted in the postmerger signal by an ultracompact horizonless object ($\epsilon \ll 1$) with different reflective properties parametrised by the shear viscosity $\eta$ of the membrane. Right panel: Ringdown of an horizonless compact object with small compactness ($\epsilon \gtrsim 0.01$) and the same reflective properties of a BH ($\eta=\eta_{\rm BH}$). The ringdown signal is modified due to the interference of the first GW echo with the prompt ringdown. Adapted from~\cite{Maggio:2020jml}.}
\label{fig:echoes}
\end{figure}

The left panel of Fig.~\ref{fig:echoes} shows the GW signal that would be emitted in the case of a horizonless compact object compared to the BH case. 
The delay time between subsequent GW echoes is fixed and depends on the width of the cavity, i.e. the compactness of the object. The delay time is computed as the round-trip time of the radiation to travel in the cavity between the light ring and the radius of the compact object. In the static and spherically symmetric case~\cite{Cardoso:2016rao},
\begin{equation}\label{delaytime}
\tau_{\rm echo} = 2 \int_{r_0}^{3M} \frac{dr}{f(r)} \sim 2 M \left[1 - 2 \epsilon -2 \log (2 \epsilon) \right] \,.
\end{equation}
The logarithmic dependence in Eq.~\eqref{delaytime} allows us to detect even Planckian corrections at the horizon scale ($\epsilon \sim l_{\rm Planck}/M$) few $\text{ms}$ after the merger with a remnant of $M\sim10M_\odot$.
The amplitude of the GW echoes depends on the reflective properties of the compact object, as shown in the left panel of Fig.~\ref{fig:echoes} for several values of the shear viscosity of the fictitious membrane. Furthermore, the light ring acts as a frequency-dependent high-pass filter, i.e. each GW echo has a lower frequency content than the previous one. At late times, the GW signal is dominated by the low-frequency QNMs of the horizonless compact object shown in Fig.~\ref{fig:lowfrequencyQNMs}.

If the remnant of a binary coalescence is a horizonless compact object with small compactness ($\epsilon \gtrsim 0.01$), the GW phenomenology in the postmerger signal would be different. In particular, the delay time of the first GW echo in Eq.~\eqref{delaytime} would be comparable with the decay time of the prompt ringdown, i.e. $\tau_{\rm ringdown} = -1/\omega_{I, \rm{BH}}\approx 10M$.
Therefore, the first GW echo would interfere with the prompt ringdown as shown in the right panel of Fig.~\ref{fig:echoes}. Finally, subsequent GW echoes are suppressed because the cavity between the light ring and the radius of the compact object is so small that does not trap the modes efficiently.

\section{Detectability}

Several searches for GW echoes have been performed based on matched-filter techniques and unmodeled searches. 
In the time domain, some phenomenological templates are based on inspiral-merger-ringdown templates in GR with additional parameters related to the morphology of GW echoes~\cite{Abedi:2016hgu} and the superposition of sine-Gaussians with free parameters~\cite{Maselli:2017tfq}. In the frequency domain, some waveform templates depend explicitly on the physical parameters of the horizonless compact object, i.e., its compactness and reflectivity~\cite{Mark:2017dnq,Testa:2018bzd,Maggio:2019zyv}.
Moreover, some unmodeled searches have been performed based on the superposition of generalized wavelets~\cite{Tsang:2018uie} and with Fourier windows~\cite{Conklin:2017lwb}.

Tentative evidence for GW echoes has been reported in the events of the first and second observing runs of LIGO and Virgo~\cite{Abedi:2016hgu,Conklin:2017lwb}, followed by independent searches arguing that the statistical significance of GW echoes is consistent with noise~\cite{Westerweck:2017hus,Nielsen:2018lkf,Tsang:2019zra,Lo:2018sep}. Furthermore, no evidene for GW echoes has been reported in the third observing run of the LIGO, Virgo, KAGRA collaboration~\cite{LIGOScientific:2021sio}. 

\begin{figure}[t]
\centering
\includegraphics[width=0.49\textwidth]{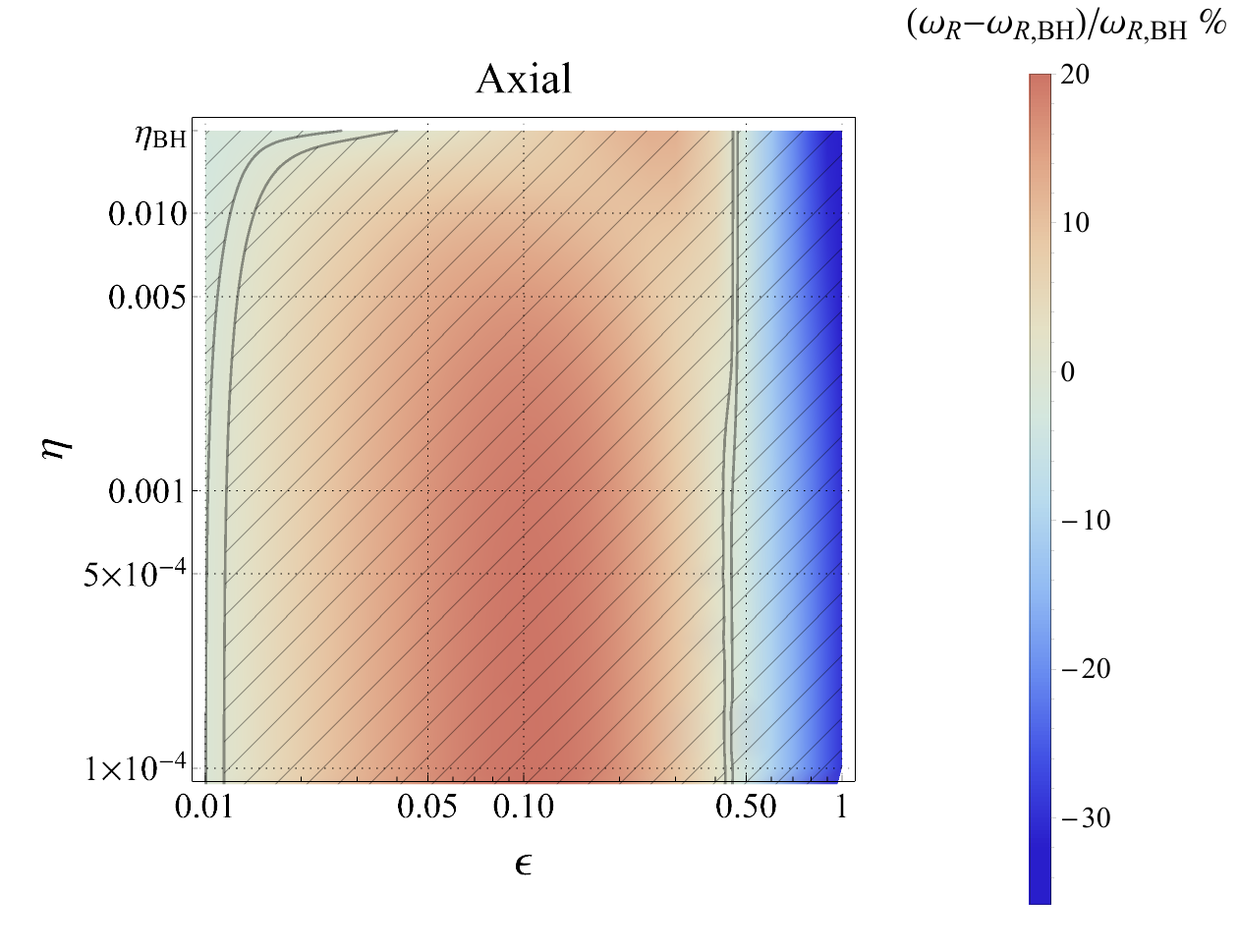}
\includegraphics[width=0.49\textwidth]{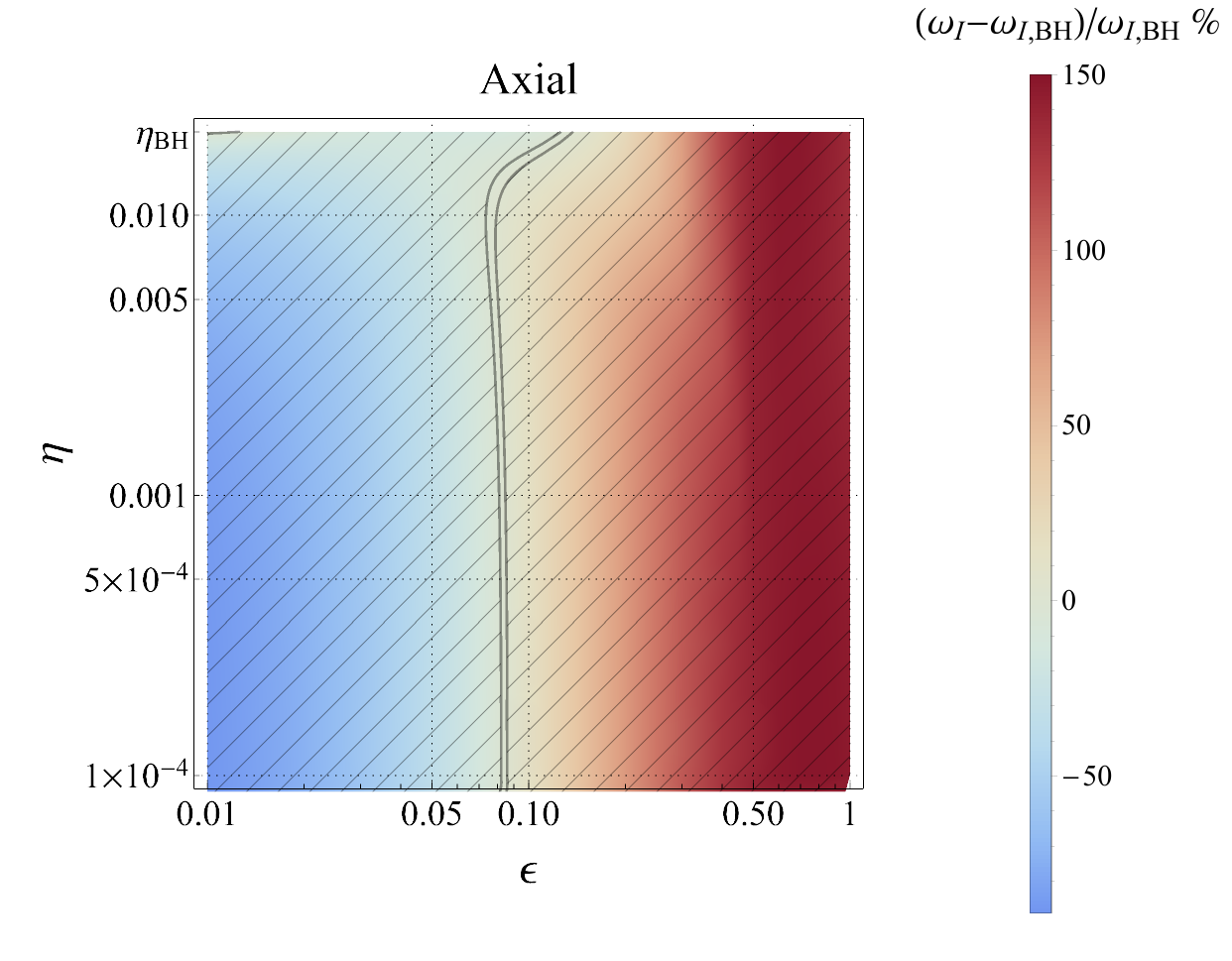}
\includegraphics[width=0.49\textwidth]{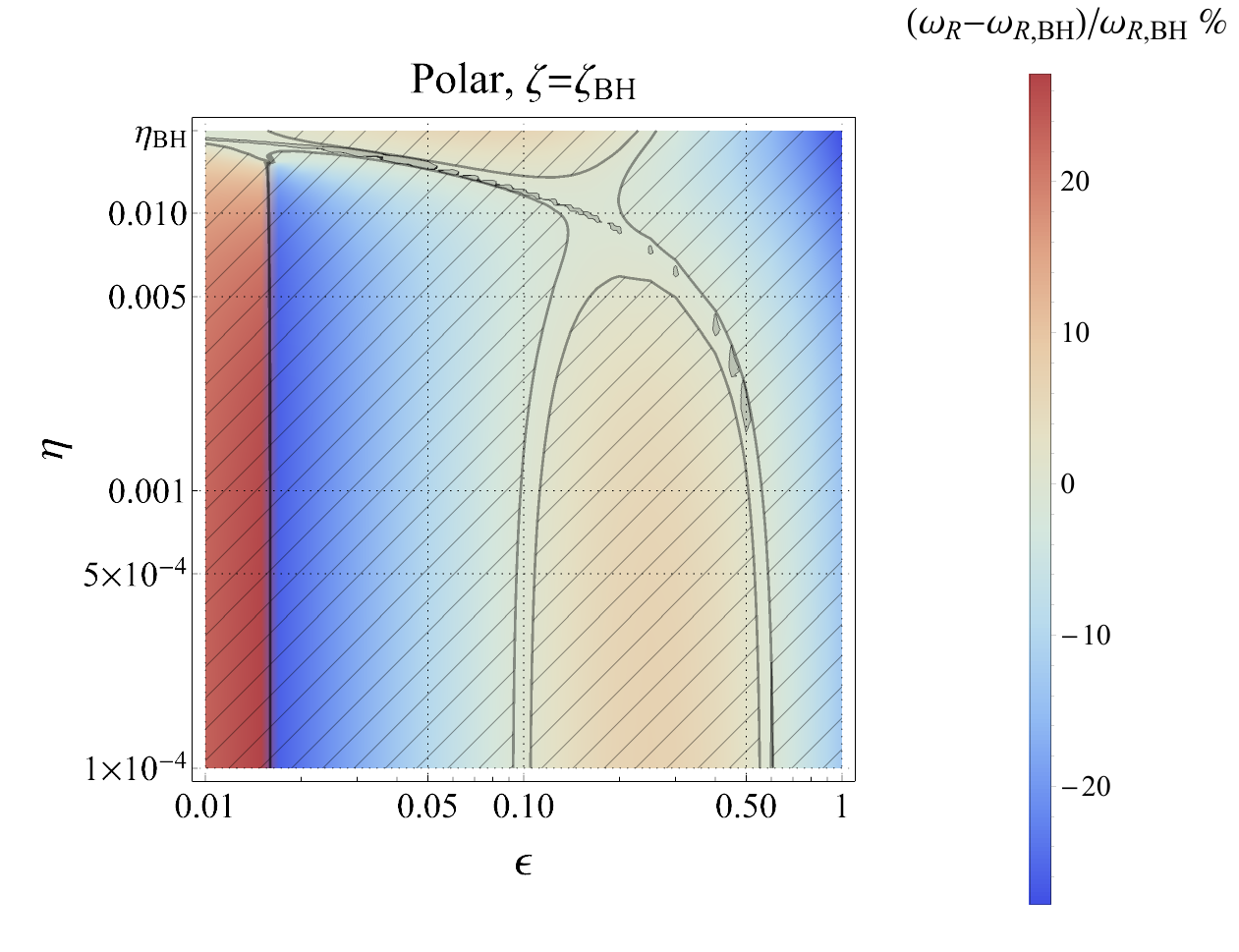}
\includegraphics[width=0.49\textwidth]{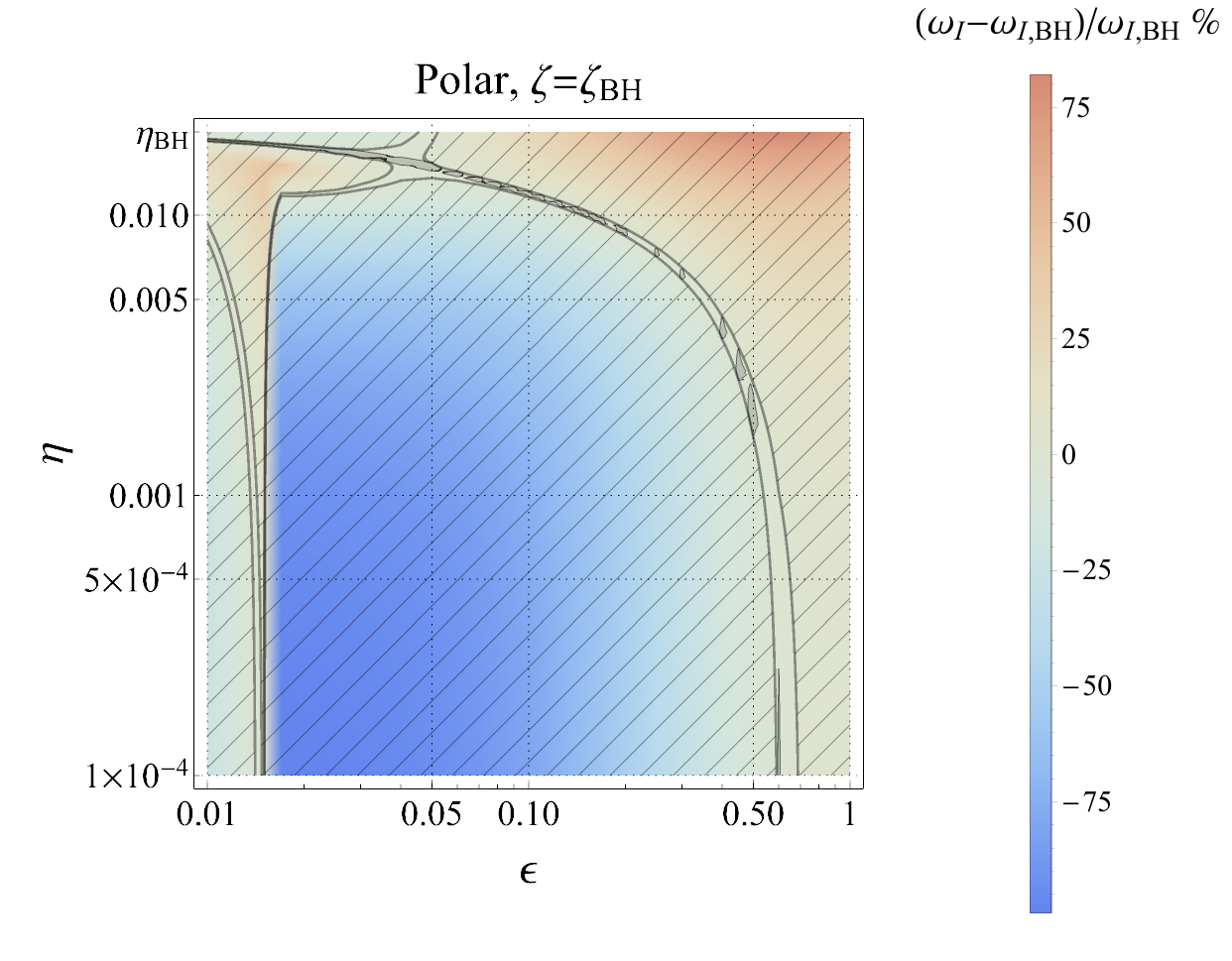}
\caption{Relative percentage difference of the real (left panels) and imaginary (right panels) part of the QNMs of a horizonless compact object to the fundamental QNM of a Schwarzschild BH for axial (top panels) and polar (bottom panels) perturbations. The dashed areas are the regions that would be excluded by individual measurements of the real and imaginary part of the QNMs by next-generation detectors.
The plot shows that next-generation detectors will allow us to constraint the whole region of the $(\epsilon, \eta)$ parameter space shown in the diagram. Adapted from~\cite{Maggio:2020jml}.}
\label{fig:plot2dSNR10}
\end{figure}
The next generation detectors have promising prospects of testing the BH paradigm. The ground-based observatories Einstein Telescope~\cite{Punturo:2010zz} and Cosmic Explorer~\cite{Reitze:2019iox} will observe GWs with an overall improvement of the signal-to-noise ratio by an order of magnitude than current detectors.
Moreover, the future space-based interferometer LISA~\cite{LISA:2017pwj} will detect GWs in the $10^{-4}-1 \ \text{Hz}$ frequency band from a variety of astrophysical sources. The sensitivity of the detectors will allow us to resolve the QNMs at percent level and perform multiple tests of the BH paradigm with the detection of higher modes.

Fig.~\ref{fig:plot2dSNR10} shows the relative percentage difference between the fundamental $\ell=2$ QNM of a Schwarzschild BH and the fundamental $\ell=2$ QNMs of a horizonless compact object with radius $r_0 = 2M(1+\epsilon)$ and reflectivity defined by the shear viscosity of the fictitious membrane. The QNM spectrum is a function of the parameter $\epsilon$ (x-axis) and the shear viscosity of the fictitious membrane $0 \leq \eta \leq \eta_{\rm BH}$ (y-axis) where $\eta=0$ describes a perfectly reflecting compact object and $\eta=\eta_{\rm BH}$ describes a totally absorbing compact object. 
The left (right) panels show the relative percentage difference of the real (imaginary) part of the QNMs for axial and polar perturbations in the top and bottom panels, respectively. 
The dashed areas are the regions of the $(\epsilon,\eta)$ parameter space that would be excluded by individual measurements of the real and imaginary part of the fundamental QNM with next-generation detectors whose accuracy is assumed to be an order of magnitude better than current detectors~\cite{Ghosh:2021mrv}.
Fig.~\ref{fig:plot2dSNR10} shows that almost the whole region of the $(\epsilon, \eta)$ parameter space would be constrained. Therefore, next-generation detectors will allow us to set very stringent constraints on the radius and the reflective properties of compact objects.

\section{Acknowledgements}
EM acknowledges funding from the Deutsche Forschungsgemeinschaft (DFG) - project number: 386119226.

%
%
%

\end{document}